\documentclass[twocolumn,showpacs,preprintnumbers,prl,superscriptaddress]{revtex4}
\usepackage[centertags]{amsmath}
\usepackage{epsfig}
\usepackage{color}
\usepackage{amsfonts}
\usepackage{graphics}
\usepackage{amsfonts}
\usepackage{amssymb}
\usepackage{amsthm}
\usepackage{newlfont}
\usepackage{dcolumn}% Align table columns on decimal point
\newcommand{\eps}{\varepsilon}

\newcommand{\bra}[1]{\left\langle#1\right\vert}
\newcommand{\ket}[1]{\left\vert #1\right\rangle}
\newcommand{\ave}[1]{\left\langle #1\right\rangle}

 \DeclareMathOperator{\im}{Im}
\DeclareMathOperator{\re}{Re}

\newcommand{\p}{\prime}

\newcommand{\fig}[1]{Fig.~\ref{#1}}
\newcommand{\wv}[3]{{}_{#1}\ave{#2}_{#3}}

\begin{document}
%%%%%%%%%%%%%%%%%%%%%%%%%%%%%%%%%%%%%%%%%%%%%%%%%%%%%%%%%%%%%%%%%%%%%%%
\title{Tomography of many-body weak values: Mach-Zehnder interferometry}

\author{Vadim Shpitalnik}
\thanks{Deceased}
\affiliation{{Department of Condensed Matter Physics, Weizmann Institute of Science, Rehovot 76100, Israel
%E-mail: Vadim.Shpitalnik@weizmann.ac.il
}}

\author{Yuval Gefen}
\affiliation{{Department of Condensed Matter Physics, Weizmann Institute of Science, Rehovot 76100, Israel
%E-mail: Yuval.Gefen@weizmann.ac.il
}}

\author{Alessandro Romito},
\affiliation{{Institut f\"ur Theoretische Festk\"orperphysik, Universit\"at Karlsruhe, D-76128 Karlsruhe, Germany}}
\affiliation{{Department of Condensed Matter Physics, Weizmann Institute of Science, Rehovot 76100, Israel
%E-mail: romito@tfp.uni-karlsruhe.de
}}

\date{\today}

%\pacs{73.23.-b}{Electronic transport in mesoscopic systems}
%\pacs{71.10.Pm}{Fermions in reduced dimensions (anyons, composite
%fermions, Luttinger liquid, etc.)}

\begin{abstract}
We propose and study a {\it weak value} (WV) protocol in the 
context of a solid state setup, specifically, an electronic Mach-Zehnder
interferometer. This is the first specific proposal to measure both the real and 
imaginary part (i.e., complete tomography) of a WV. We also analyze the manifestation
of many-body physics in the WV to be measured, including finite temperature
and shot-noise-like contributions.
\end{abstract}


\maketitle
%%%%%%%%%%%%%%%%%%%%%%%%%%%%%%%%%%%%%%%%%%%%%%%%%%%%%%%%%%%%%%%%%%%%%%
%Introduction

The problem of non-invasive measurement of a quantum system is
interesting both from the view point of foundations of quantum
mechanics and possible applications (e.g. quantum computation). As
opposed to the standard strong measurement procedure described by
the projection postulate \cite{Neuman}, weak measurement of an
observable, while weakly disturbing the system, provides only
partial information on the state of the latter. Normally the
outcome of a weak measurement of an observable is its expectation
value, which is the weighted average of the eigenvalues of this
observable. Major step in formulating alternatives to standard
measurement procedure was the proposal of a {\it weak value}
protocol~\cite{Aharonov}, consisting of a weak
measurement (of $\hat{A}$), followed by a strong one (of
$\hat{B}$). The outcome of the first is conditional on the
result of the second (post-selection). Specifically, if the system
is prepared (preselected) in the state $\ket{\chi_i}$, and is
postselected in the state $\ket{\chi_f}$, the weak value of the
operator $\hat{A}$ is
\begin{equation}\label{qm_wv_def}
\wv{\chi_f}
{A}{\chi_i}=\frac{\langle\chi_f|\hat{A}|\chi_i\rangle}{\langle\chi_f|\chi_i\rangle}
=\frac{\langle\chi_i|\Pi_{\chi_f}\hat{A}|\chi_i\rangle}{\langle\chi_i|\Pi_{\chi_f}|\chi_i\rangle},
\end{equation}
where $\Pi_{\chi_f}=\ket{\chi_f}\bra{\chi_f}$ is a projection operator.

In general, the weak value is a complex number and (as opposed to a
projective expectation value) its real part can be
out of the range of the eigenvalues of $\hat{A}$. In order to
obtain non-trivial weak values (e.g., outside the range of the eigenvalues of $\hat{A}$; 
complex; negative when $\hat{A}$ is positive definite), the preselected state should not be
an eigenstate of the measured operator, and $[\hat{A},\hat{B}] \neq 0$.

WV may allow us to explore some fundamental aspects of
quantum measurement, including access to simultaneous measurement
of non-commuting variables~\cite{Hongduo:2007aa,Jordan:2005aa};
dephasing and phase recovery~\cite{Neder:2007a};
correlation between measurements~\cite{Sukhorukov:2007aa}; and even
new horizons in metrology~\cite{Aharonov,Hosten:2008}. While some aspects
of WVs have been demonstrated in optics based
setups~\cite{Pryde:2005}, the
implementation of weak values in the context of solid state
physics is very new~\cite{Romito:2008, Williams:2007}.

Weak values are not just an artifact of a convoluted definition,
but indeed do emerge as the outcome of weak measurement of a pre-
and post-selected states. Microscopically 
we consider the coupling of
a system to a detector with a Hamiltonian $H=H_{
\textrm{S}}+H_{\textrm{D}}+H_{\textrm{int}}$. The interaction Hamiltonian is 
$H_{\textrm{int}}=\lambda g(t) \hat{p}\hat{A}$.
Here $\hat{p}$ is the momentum canonically conjugate to the
position of the detector's pointer, $\hat{q}$, and
$\lambda g(t)$ ($\lambda\ll1$) is a time dependent coupling
constant~\cite{foot1}. Following the weak measurement and the
post-selection steps, the expectation value of the coordinate of the
pointer (initially equal to $q_0$) is given by
$\langle \hat{q}\rangle=q_0-\lambda \re[_{\psi}\langle
A\rangle_{\phi}]$, i.e., the shift in the pointer is proportional to the real part of
the WV. Under more general conditions (i.e., $\hat{p}$, $\hat{q}$ are not canonically 
conjugated)the imaginary part of the WV
may be  meaningful too~\cite{Steinberg:1995}. An experimental
procedure which will provide for a full "tomography" of (both the
real and imaginary parts of) the weak value remains a challenge.

\begin{figure}
\begin{center}
\includegraphics[width=85mm]{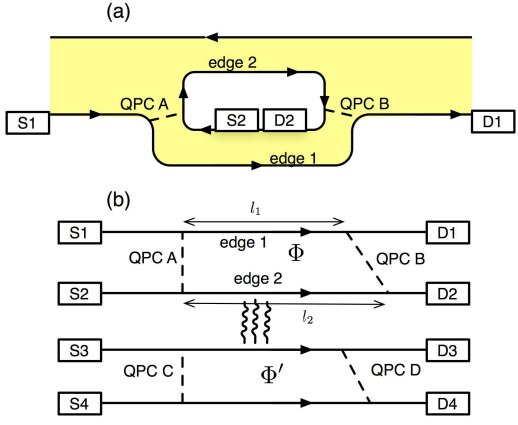}
\end{center}
\caption{
(a) Electronic MZI represented by edge states (full lines)
in a Hall bar. Inter-edge tunneling (dashed lines) takes place at
the quantum point contacts. 
(b) Scheme of a double electronic MZI. Electrons on
edge states 2 and 3 interact via Coulomb interaction.
 } \label{MZI}
\end{figure}

Here we report the first  systematic study 
of {\it complex} WVs in the context of {\it many electron} 
solid state system,
 proposing an experimental procedure which will provide for the 
full ``tomography'' of (both the real and imaginary part of)
WVs.
Employing building blocks which are accessible within current
technology~\cite{Ji:2003}, our proposed
protocol and the ensuing predictions are amenable to experimental
test. In particular, addressing a ``system'' and a ``detector'' 
which are represented by an electronic Mach-Zehnder interferometer
(MZI)~\cite{Ji:2003} (\fig{MZI}), (i) We propose
how to retrieve both the real and the imaginary parts; 
(ii) We show that the introduction of both a non-pure
state and finite temperature ($T$) 
modify the WV and reduce the
visibility of Aharonov-Bohm (AB) oscillations; (iii) We show
how many-body effects lead to reduction of the weak value's visibility, as a
function of voltage bias. 
 In particular $\mbox{\rm Im}[\wv{\chi_f}
{A}{\chi_i}]$ maintains its single particle value while 
$\mbox{\rm Re}[\wv{\chi_f}{A}{\chi_i}]$ is modified by an excess 
(non-equilibrium) noise and thermal noise terms;
(iv) We show that for the system at hand the WV is related to a system-detector
current-current cross correlator~\cite{foot2}.

The chiral leads of the MZI are realized by the edge states of
an integer quantum Hall setup. Throughout this analysis we will
consider non-interacting single edge channels ($\nu=1$)~\cite{foot6}.  
We do account here for the system-detector
interaction and for the inherent many-fermion physics.

%%%%%%%%%%%%%%%%%%%%%%%%%%%%%%%%%%%%%%%%%%%%%%%%%%%%%%%%%%%%%%%%%%%
%%%%%%%%%%%%%%%%%%%%%%%%%%%%%%%%%%%%%%%%%%%%%%%%%%%%%%%%%%%%%%%%%%%

%%%%%%%%%%%%%%%%%%%%%%%%%%%%%%%%%%%%%%%%%%%%%%%%%%%%%%%%%%%%%%%%%%%%%%%
%\section{The system}
Our model Hamiltonian for the system (S) and the 
detector (D) is~\cite{Chalker:2007}
\begin{eqnarray}
& &H_S=-i v\sum_{m=1}^2\int
dx:\psi^{\dag}_m(x)\partial_x\psi_m(x): \nonumber \\
& &
+\Gamma_a\psi^{\dag}_1(0)\psi_2(0)+\Gamma_b\psi^{\dag}_1(l_1)\psi_2(l_2)
+H.c.\, , \\
& &H_D=-i v\sum_{m=3}^4\int
dx:\psi^{\dag}_m(x)\partial_x\psi_m(x):\nonumber\\
& &+
\Gamma_c\psi^{\dag}_3(0)\psi_4(0)+\Gamma_d\psi^{\dag}_3(l_3)\psi_4(l_4)
+H.c.
\end{eqnarray}
Here $\psi^{\dag}_m(x)=\frac{1}{\sqrt{L}}\sum_{k=-\infty}^{\infty}
e^{ikx}d^{\dag}_{km}$ is the creation operator at point $x$ on
arm $m$ (the system's length, $L$, is assumed to be much
larger than any other length scale in the problem). The $\Gamma$'s are
inter-edge tunneling amplitudes, 
normal ordering is with respect to the vacuum (states are
occupied for $k\leq0$ and empty otherwise), 
$v$ is the edge velocity and $\hbar=1$. 
We assume a weak capacitive interaction between
the system and the detector:
\begin{displaymath}
H_{\textrm{int}}=\int dx \, dy  e^2 U(x,y) :\psi^{\dag}_2(x)\psi_2(x):
:\psi^{\dag}_3(y)\psi_3(y): \, ,
\end{displaymath}
where the support of $U(x,y)$ is $0<x,y<\min_m{(l_m)}$. Without loss of
generality we assume that sources $S1$
and $S3$ are voltage biased ($V$ and $V_3$), 
and use the current at $D4$ as the weak measurement's pointer.

$H_{S}$ is readily diagonalized.
In the Heisenberg representation the system's $\psi$ 
operators are  
\begin{gather}\label{solution}
\psi_m(x,t)=\frac{1}{\sqrt{L}}\sum_ke^{ik(x-vt)} \left\{
\begin{array}{ll}
     a_{km}, & \hbox{$x<0;$} \\
    b_{km}, & \hbox{$0<x<l_m;$} \\
    c_{km}, & \hbox{$l_m<x.$} \\
\end{array}
\right.
\end{gather}
where $m=1,2$ and $\mathbf{b}_k=\mathcal{S}_A \mathbf{a}_k$,
$\mathbf{c}_k=\mathcal{S}_B \mathbf{b}_k$ with scattering matrices
$ \mathcal{S}_i = \left(%
\begin{array}{cc}
  r_i & t_i \\
  -t_i^* & r_i \\
\end{array}%
\right)$ ($i=a,b$) and, for equal arm lengths, 
$r_i=[(2v)^2-|\Gamma_i|^2]/[(2v)^2+|\Gamma_i|^2]$,  
$t_i=4iv\Gamma_i/[(2v)^2+|\Gamma_i|^2]$.
The effect of the system's AB phase and unequal arm's lengths can be 
absorbed into  $\tilde{\Gamma}_A \equiv \Gamma_A e^{i\tilde{\phi}/2}$,
$\Gamma_B \equiv \Gamma_B e^{-i\tilde{\phi}/2}$. Here
$\tilde{\phi}$ includes the contributions of both the AB flux, $\Phi$, 
and the orbital phase: $\tilde{\phi} \equiv \phi+k(l_2-l_1)$, 
$\phi \equiv 2 \pi \frac{\Phi}{\Phi_0}$
 ($\Gamma_A, \Gamma_B >0$). 
 A similar expression is available for $H_D$ as well.

%%%%%%%%%%%%%%%%%%%%%%%%%%%%%%%%%%%%%%%%%%%%%%%%%%%%%%%%%%%%%%%%%%%%%%%
%\section{Weak values in the electronic MZI}

As a first step we consider the injection of a {\it single particle} (in
the scattering state $k$) into the system~\cite{foot3} (e.g., at the source S1). 
The preselected state is then $\ket{S_1}=a_{k1}^{\dag}\ket{0}$. 
The post-selected state is $\ket{D_2}=c_{k2}^{\dag}\ket{0}$
(detection at D2). By definition \eqref{qm_wv_def} the WV of
the charge at the intermediate segment of arm $i$ is
$_{D2}\langle \hat{Q}_i\rangle_{S1}^{\textrm{SP}}/e=  \frac{\langle
c_{k2}b^{\dag}_{ki}b_{ki}a^{\dag}_{k1}\rangle}{\langle
c_{k2}a^{\dag}_{k1}\rangle}$,  yielding 
$_{D2}\langle \hat{Q}_2\rangle_{S1}^{\textrm{SP}}/e
= \frac{t_A^* r_B}{t_A^* r_B + e^{i\tilde{\phi}} r_A t_B^*}$ (pure state).
This quantity may assume complex values depending on the choice
of parameters (transmission and reflection amplitudes and $\Phi$).
In particular,  its real part may be larger than $1$ or even negative, 
clearly outside the allowed range  of the number operator expectation value. 
But, as expected, $_{D2}\langle Q_1 \rangle_{S1}^{\textrm{SP}}+
 _{D2}\langle Q_2 \rangle_{S1}^{\textrm{SP}}=e$. 
The individual weak values are $\Phi_0$ periodic with the flux.

In a more general case, when the system is initially not in 
a pure state (e.g., finite temperature, entanglement, etc.), eq.~(\ref{qm_wv_def})
is not applicable, and we have to rewrite it in terms of the density matrix,
 $\rho$, of the initial state, yielding~\cite{Romito:2008}:
\begin{equation}
\wv{\chi_f}{A}{\chi_i} \equiv \frac{\mathop{Tr}(\Pi_{\chi_f}\hat{A}\rho)}{\mathop{Tr}(\Pi_{\chi_f}\rho)}
= \frac{\langle \chi_f |\hat{A} \rho| \chi_f \rangle}{\langle\chi_f |\rho| \chi_f \rangle} \, . \nonumber
\end{equation} 
For example assume that a voltage bias $V$ is applied to the source S1,
which is held at temperature $T$. Then $\rho=1/L 
\sum_k [f(vk-eV)-f(vk)]a_{k1}^{\dag}a_{k1}$, with $f(x)$ the 
Fermi-Dirac distribution at zero voltage bias, and
\begin{equation}
_{D2}\langle Q_2 \rangle_{S1}^{\textrm{SP}} =
\frac{t_A^* r_B(t_A r_B + K(V,T)e^{-i\varphi} r_A t_B)}{|r_A t_B|^2+|t_A r_B|^2 + 2 K(V,T)\mbox{\rm{Re}} [e^{i\varphi} r_A t_B^* t_A r_B]} \, ,
\label{mixed_state}
\end{equation}
 where $\varphi=eV\Delta l / (2v) +\phi$, $K(V,T)\equiv \frac{2\pi k_B T \sin (eV \Delta l/(2v))}{eV\sinh(\pi k_B T \Delta l /v) }$
 and $\Delta l=l_2-l_1$. We observe that the broadening of the 
 energy and non-pure nature of the initial state lead to the suppression of the 
 interference term, which in turn leads to the suppression of the imaginary part
 of the WV and reduces the real part to a "standard" value lying in the interval $[0,1]$.
 
We next study the generalization of this protocol to 
a {\it many-body} (albeit non-interacting) system.
The preselected state is given by the stationary voltage biased ($V$)
non-equilibrium state of the system. Post-selection will be
defined by the detection of {\it one} electron at the drain D2, i.e.
within  $[x-\eps,x+\eps]$, $\epsilon \rightarrow 0$ around $x>l_1,l_2$. The
projection operator for this state is $:\psi_2^{\dag}(x)\psi_2(x):$.
For a system with linear dispersion relation, the
current operator is $\hat{I}_2(x,t)=ev:\psi_2^{\dag}(x,t)\psi_2(x,t):$; 
it follows that
we can employ $\hat{I}_2/(ev)$ as a projection operator for the post-selection. 
We consider the WV of charge (or current) at the intermediate segment
of arm $2$. It is given by the operator
$\hat{Q}_2(x_0)=e v \int_{-\tau/2}^{\tau/2}dt:\psi^{\dag}_2(x_0,t)\psi_2(x_0,t):$,
 where $\tau$ is an infrared cut-off, $\tau\gg\frac{L}{v}$, and
$0<x_0<l_2$. Such an operator can be measured e.g. by adiabatically
switching on (off) the coupling to the measurement device at 
$t<0$ ($t>0$)~\cite{foot4}. 
From \eqref{qm_wv_def}, and employing $\langle
a^{\dag}_{ki}a_{ki}\rangle=f(vk-\mu_i)$ with
$\mu_1=eV,~\mu_2=0$, we obtain~\cite{foot5}:
\begin{widetext}
\begin{equation}
\label{mb_weak_val}
\wv{D2}{Q_2}{S1}^{\textrm{MB}}  =  \frac{\langle I_2(x,0)Q_2(x_0)\rangle}
{\langle I_2(x,0)\rangle} = \frac{e}{1/L \sum_k (f_{eV}-f_0) \mathcal{A}_{11}}\Bigl[\sum_{i,j=1,2}
\frac{1}{L}\sum_k \mathcal{A}_{ij}\mathcal{B}_{ji}f_{\mu_i}(1-f_{\mu_j})\Bigr] + v\tau \sum_k (f_{eV}-f_0)\mathcal{B}_{11}\Bigr]\, .
\end{equation}
\end{widetext}
Here $\mathcal{A}_{ij} \equiv (\mathcal{S}_A^{\dag}\mathcal{S}_B^{\dag})_{i2}
(\mathcal{S}_B\mathcal{S}_A)_{2j}$, $\mathcal{B}_{ij}\equiv
(\mathcal{S}_A^{\dag})_{i2}(\mathcal{S}_A)_{2j}$ and $f_{\mu}$ 
stands for the Fermi distribution function $f(vk-\mu)$.
%\begin{widetext}
%\begin{eqnarray}
%= e^2v^2\int_{-\tau/2}^{\tau/2}dt 
%\frac{\langle[\psi^{\dag}_2(x,0)\psi_2(x,0)-\rho_0(x)]
%[\psi^{\dag}_2(x_0,t)\psi_2(x_0,t)-\rho_0(x_0)]\rangle}
%{\langle I_2(x,0)\rangle} 
%= \frac{e^2v^2}{L^2\langle I_2(x,0)\rangle} \nonumber \\
%& & \times \int_{-\tau/2}^{\tau/2}dt  \sum_{k_1,k_2}\sum_{i,j=1,2}\Bigl[\mathcal{A}_{ij}
%\mathcal{B}_{ji}e^{i(k_2-k_1)((x-x_0)-vt)}
%\langle a^{\dag}_{k_1i}a_{k_1i}\rangle\langle
%a_{k_2j}a^{\dag}_{k_2j}\rangle 
%+\mathcal{A}_{ii}\mathcal{B}_{jj}(\langle
%a^{\dag}_{k_1i}a_{k_1i}\rangle - f(vk_1))(\langle
%a^{\dag}_{k_2j}a_{k_2j}\rangle - f(vk_2))\Bigr]\, .
%\end{eqnarray}
%\end{widetext}
%Here $\rho_0(x)$ is the average density at equilibrium,
% and $\mathcal{A}_{ij}\equiv
%(\mathcal{S}_A^{\dag}\mathcal{S}_B^{\dag})_{i2}
%(\mathcal{S}_B\mathcal{S}_A)_{2j}$, $\mathcal{B}_{ij}\equiv
%(\mathcal{S}_A^{\dag})_{i2}(\mathcal{S}_B)_{2j}$. Employing $\langle
%a^{\dag}_{ki}a_{ki}\rangle=f(vk-\mu_i)$ with
%$\mu_1=eV,~\mu_2=0$ and $\delta n \equiv \frac{1}{L}\sum_k(f(vk-eV) -
%f(vk))$ we obtain~\cite{foot5}
%\begin{widetext}
%\begin{gather}\label{mb_weak_val}
%\wv{D2}{Q_2}{S1}^{\textrm{MB}}=
%\frac{e}{v\delta n \mathcal{A}_{11}}\Bigl[\sum_{i,j=1,2}\mathcal{A}_{ij}\mathcal{B}_{ji}\frac{1}{L}\sum_k
%f(vk-\mu_i)[1-f(vk-\mu_j)] + \mathcal{A}_{11}\mathcal{B}_{11}v\tau (\delta n)^2\Bigr]\, .
%\end{gather}
%\end{widetext}
The summation over $k$  is replaced by energy integrals,
$1/L\sum_k \rightarrow 1/(2\pi v) \int dE$.
At zero temperature the WV \eqref{mb_weak_val} is:
\begin{gather}\label{zero_temp_wv}
\wv{D2}{Q_2}{S1}^{\textrm{MB}}=
 \wv{D2}{Q_2}{S1}^{\textrm{SP}}+(N-1)  \langle Q_2 \rangle  \,\,\,\, (T=0) \,  .
\end{gather}
Here $N=\frac{eV\tau}{2\pi}$ is the number of electrons
injected into the MZI during the measurement time. 
The first term on the r.h.s is the single
particle WV \eqref{mixed_state}. 
The second term is proportional to the single
particle expectation value of $\hat{Q}_2=|t_A|^2$. This can be interpreted
in the following way: the post-selection enforces a detection of
one electron at point $x$ at time $t$. The correlation between
$\hat{Q}_2$ and a post-selection yields the first term. All other
electrons which are injected into the system during the
measurement are not constrained by the postselection and therefore
give rise to an additional contribution to $\wv{D2}{Q_2}{S1}^{\textrm{MB}}$.

%%%%%%%%%%%%%%%%%%%%%%%%%%%%%%%%%%%%%%%%%%%%%%%%%%%%%%%%%%%%%%%%%%%
\begin{figure}
\begin{center}
\includegraphics[width=85mm]{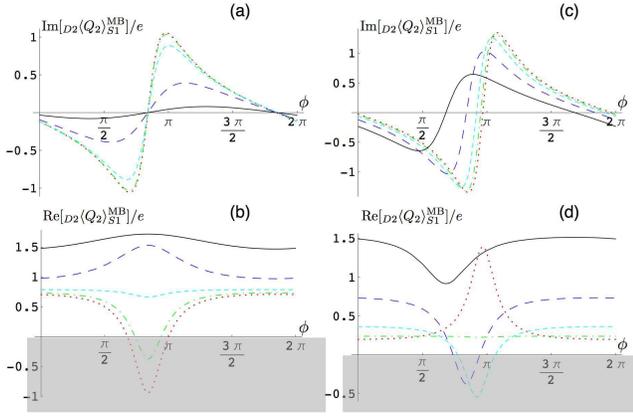}
\end{center}
\caption{Real and Imaginary part of the many-body WV as a function of 
the Aharanov-Bohm phase.  Panels (a), (b)  correspond to 
different plots for different 
values of temperature: $T \Delta/l/(\hbar v)=$$0$ (dotted), $0.05$ (dash-dotted), 
$0.15$ (dashed), $0.5$ (long-dashed), $1.2$ (full). $V =5\mu e$V. 
Panels (c), (d) correspond to 
different plots for different voltage biases: 
$V \Delta l/(\hbar v)=$$0.2$ (dotted), $0.3$ (dash-dotted), $0.5$ (dashed), 
$1$ (long-dashed), $2$ (full). $T=3$mK. 
In all plots $\hbar v / \Delta l=5 \times 10^{-6}e$V,
$\tau=2 \times 10^{-10}$ $s$, $t_A=r_A=1/\sqrt{2}$, $t_B=\sqrt{2}r_B=\sqrt{2}/\sqrt{3}$. 
The range for which $\mbox{\rm{Re}}[_{D2}\langle Q_2 \rangle_{S1}^{\textrm{SP}}]$
is beyond the interval of standard (strong) values is shadowed.} \label{risultati}
\end{figure}
%%%%%%%%%%%%%%%%%%%%%%%%%%%%%%%%%%%%%%%%%%%%%%%%%%%%%%%%%%%%%%%%%%%

Consider now a finite temperature. Using $ \frac{1}{L}\sum_k
f_i(k)(1-f_i(k))\simeq-k_BT\int\frac{dE}{2\pi v}\frac{\partial
f}{\partial E}$ we can write  the weak value as
\begin{widetext}
\begin{eqnarray}
 \frac{\wv{D2}{Q_2}{S1}^{\textrm{MB}}}{e} &= & \mbox{\rm{Re}} \left[ 
\frac{ F(V,T) \left( | r_B r_A t_A|^2 - |t_B t_A r_A|^2 \right) + \left( |r_A|^2 - |t_A|^2 \right)  
\bar{K}(V,T) e^{i \varphi} r_A t_A r_B t_B^*}{| r_A t_B |^2 + | t_A r_B |^2 + 2 K(V,T) \mbox{\rm{Re}}( e^{i \varphi}t_A r_B r_A t_B^*)} +|t_A|^2 \right] \nonumber \\
 & & + i \mbox{\rm{Im}}[\frac{\wv{D2}{Q_2}{S1}^{\textrm{SP}}}{e}]
+  \frac{k_B T |r_B|^2}{eV \left(| r_A t_B |^2 + | t_A r_B |^2 + 2 K(V,T) \mbox{\rm{Re}}( e^{i \varphi}t_A r_B r_A t_B^*) \right)} 
+ \langle Q_2 \rangle (N-1).
\end{eqnarray}
\end{widetext}
Here $F(V,T)=\coth[eV/(2k_B T)]-2k_B T/(eV)$, $K(V,T)$ 
is given by Eq.~(\ref{mixed_state}), and~\cite{Chung:2005}
$\bar{K}(V,T)=2\pi k_B T
[ \sin(eV \Delta l / 2v) \coth( eV / 2k_B T)- k_B T \Delta l / v \cos( eV \Delta l / 2v)] 
/ [eV\sinh( \pi k_B T \Delta l / v )]$. 
This is the central result of the letter. 
The many body weak value in electronic MZI consists 
of four terms: 
(i) A term which is related to the real part of
the single particle weak value corrected by finite
temperature Fermi sea effects;
(ii) The imaginary part of the single particle weak value; 
(iii) A contribution due to thermal fluctuations~\cite{Martin:1992}; 
(iv) Many body ``excess noise''. 
The dependence of the many-body WV on voltage and temperature is 
presented in \fig{risultati}.
Note the asymmetry between the imaginary and the real parts of the WV:
the former coincides with the 
imaginary part of the single particle weak independently on the voltage 
bias and temperature.
When the temperature is smaller than the voltage bias, $k_B T \ll eV$ 
the many-body VW reduces to the expression in Eq.(\ref{zero_temp_wv}),
while at high temperature the third term dominates and thermal fluctuations 
wash out the SP weak values.
%\begin{gather}
%\wv{D2}{Q_2}{S1}^{\textrm{MB}} \approx
%e \frac{k_BT}{eV}\frac{|r_B|^2}{|r_A t_B + t_A r_B|^2} \, .
%\end{gather}

%%%%%%%%%%%%%%%%%%%%%%%%%%%%%%%%%%%%%%%%%%%%%%%%%%%%%%%%%%%%%%%%%%%%%%%
%\section{The measurement and tomography of weak values}
We now describe the measurement procedure leading to the WVs.
We should calculate the expectation value of  
the pointer's variable, which, in our case, is the time average 
of the outgoing current in the drain D4. The latter is given by 
the expectation value of 
$\hat{I}_4(x^{\p})=1/\tau \int_{-\tau/2}^{\tau/2} dt\, \hat{I}_4 (x',t)$
($x^{\p}>l_4$), in the detector's state after postselection.
 Post-selection is realized by acting on the
system's state with the projection operator $\hat{I}_2(x,t)$. Therefore
the effective action of the detector after performing the
post-selection and tracing over the system's state is
\begin{equation}
e^{iS_{D}^{(\textrm{eff})}}=\int \mathcal{D}[\bar{\psi}_1 \psi_1] \mathcal{D}[\bar{\psi}_2 \psi_2] \, I_2(x,t)e^{iS},
\end{equation}
and the effective partition function is $Z_{D}^{(\textrm{eff})}= \int
\mathcal{D}[\bar{\psi}_3\psi_3]\mathcal{D}[\bar{\psi}_4\psi_4]\, 
e^{iS_{D}^{(\textrm{eff})}}$. We calculate the
expectation value of $\hat{I}_4(x^{\p},t)$ using this effective
partition function, which yields $\ave{\hat{I}_4(x^{\p})}=1/Z_{D}^{(\textrm{eff})}
\int\mathcal{D}[\bar{\psi}_3\psi_3]\mathcal{D}[\bar{\psi}_4\psi_4]\, 
I_4(x^{\p})e^{iS_{D}^{(\textrm{eff})}} =
\ave{\hat{I}_4(x^{\p})\hat{I}_2(x,t)} / \ave{\hat{I}_2(x,t)}$.
Hence, in order to
obtain a weak value we consider the correlator
$\ave{I_2(t)I_4(t^{\p})}$. Since the system is out of equilibrium
we employ the Keldysh technique to calculate this correlator. 
After some algebra we obtain to first order in the interaction
\begin{gather}
\frac{1}{\tau}\int_{t-\tau/2}^{t+\tau/2}d t^{\p} \, \ave{I_2(t)I_4(t^{\p})}
 =\ave{I_2}_0 \langle  I_4\rangle_0 \cr
\times  \Bigl(1+\frac{2\lambda}{e^2\tau}\im\Bigl\{(_{D2}\ave{Q_2}_{S1}^{\textrm{MB}})
(_{D4}\ave{Q_2}_{S3}^{\textrm{MB}})\Bigr\}\Bigr)\, ,
\end{gather}
where $\lambda=e^2\int dx \, dy \, U(x, y)$, and $\ave{\phantom{i}}_0$ refers
to the $\lambda=0$ expectation value. 
Indeed , dividing by $\ave{I_2}$, we see that the 
shift in the detector's current is proportional to the 
WV~\eqref{mb_weak_val}. Since here the ``system'' and the ``detector''
play a symmetric role, the result can be stated in a 
system-detector symmetric fashion: 
for weak enough interaction,
the current-current correlator is proportional to the product of
WV of the respective system and detector operators which appear 
in $H_{\textrm{int}}$.
Notice that varying the Aharonov-Bohm flux in the "detector" MZI
one can explore the real and the imaginary parts of the "system's"
WV.

The analysis presented here is the first significant step towards 
a complete characterization of weak values in interacting systems.
The visibility of flux sensitive many-body WVs
 is reduced by both $T$ and $V$: 
The single particle WV is supplemented by thermal noise 
and by a voltage dependent many fermion term. This WV tomography
and the tuning of the respective noise terms are all amenable 
to experimental verification.

%Finally we comment on the effect of the asymmetry. As is shown at
%\cite{Chung:2005}, the asymmetry of the interferometer leads to
%the suppression of the visibility of AB oscillations as a function
%of temperature and bias. Specifically, in order to account for the
%asymmetries, in all above calculations in the interference terms
%we have to change
%\begin{displaymath}
%\cos(\phi)\rightarrow 2\pi k_bT\frac{\sin(eV\Delta L/2\hbar
%v)}{\sinh(\pi k_bT\Delta L/\hbar v)}\cos\Bigl(\phi+ \frac{eV\Delta
%L}{2\hbar v}\Bigr)
%\end{displaymath}

%%%%%%%%%%%%%%%%%%%%%%%%%%%%%%%%%%%%%%%%%%%%%%%%%%%%%%%%%%%%%%%%%%%%%%
%\section{Acknowledgements}
We acknowledge illuminating discussions with Y. Aharonov, 
A. Stern, and L. Vaidmann. This work was supported by
the Minerva Foundation, US-Israel BSF and the DFG project SPP
1285, and DFG Priority Programme "Semiconductor Spintronics".

%%%%%%%%%%%%%%%%%%%%%%%%%%%%%%%%%%%%%%%%%%%%%%%%%%%%%%%%%%%%%%%%%%%%%%%

%%%%%%%%%%%%%%%%%%%%%%%%%%%%%%%%%%%%%%%%%%%%%%%%%%%%%%%%%%%%%%%%%%%%%%%%%%%%%%%%%%%%%%%%%%%%%%%%%%%%%%%%%%%%%%5


\begin{thebibliography}{10}
\bibitem{foot1}
In this example we have assumed that the free Hamiltonians of the
system and the detector vanish and that $g(t)=\delta(t-t_0)$.
\bibitem{foot2}
We believe that a qualitatively similar statement 
(concerning a system-detector two operator cross-correlator)
applies to weak value protocols at large.
\bibitem{foot3}
Note that a single particle MZI can be mapped onto the problem of spin $1/2$ particle. 
Indeed, choose the $z$-axis such that $\ket{\uparrow}_z=b_1^{\dag} \ket{0}$ and
$\ket{\downarrow}_z=b_2^{\dag} \ket{0}$. Then $\sigma_z=b_1^{\dag} b_1 -b_2^{\dag} b_2$,
the scattering matrices rotate the spin in, say, the $xz$ plane and the AB flux rotate the spin around the $z$-axis.
\bibitem{foot5}
This expression is readily generalized to the situation in which bias is applied to 
the source S$_{\alpha}$, the post-selected electron is detected at D$_{\beta}$, and the 
weakly measured operator is $Q_{\gamma}$. In this case one has to replace the index
$2$ in the definition of $\mathcal{A}_{ij}$ by $\beta$, and in $\mathcal{B}_{ij}$ by $\gamma$. 
The index $1$ in $f_1$, $\mathcal{A}_{11}$, and $\mathcal{B}_{11}$ is replaced by 
the index $\alpha$.
\end{thebibliography}
\end{document}